\documentclass[preprint]{aastex}
\usepackage{natbib}

\slugcomment{Accepted by the \emph{Astrophysical Journal}}

\newcommand{\WMAP}{{\sl WMAP}}

\received{2003 February 11}
\begin{document}

\title{First Year {\sl Wilkinson Microwave Anisotropy Probe} (\WMAP) 
       Observations: Beam Profiles and Window Functions}

\author{
L. Page\altaffilmark{2},
C. Barnes\altaffilmark{2},
G. Hinshaw\altaffilmark{3},
D. N. Spergel \altaffilmark{4},
J. L. Weiland \altaffilmark{5},
E. Wollack \altaffilmark{3},
C. L. Bennett\altaffilmark{3},
M. Halpern \altaffilmark{6},
N. Jarosik \altaffilmark{1}, 
A. Kogut \altaffilmark{2}, 
M. Limon \altaffilmark{2,7}, 
S. S. Meyer \altaffilmark{8},
G. S. Tucker \altaffilmark{7,9}, 
E. L. Wright \altaffilmark{10}
}

\altaffiltext{1}{\WMAP\ is the result of a partnership between Princeton 
                 University and NASA's Goddard Space Flight Center. Scientific 
                 guidance is provided by the \WMAP\ Science Team.}
\altaffiltext{2}{Dept. of Physics, Jadwin Hall, 
            Princeton, NJ 08544}
\altaffiltext{3}{Code 685, Goddard Space Flight Center, 
            Greenbelt, MD 20771}
\altaffiltext{4}{Dept of Astrophysical Sciences, 
            Princeton University, Princeton, NJ 08544}
\altaffiltext{5}{Science Systems and Applications, Inc. (SSAI), 
            10210 Greenbelt Road, Suite 600 Lanham, Maryland 20706}
\altaffiltext{6}{Dept. of Physics and Astronomy, University of 
            British Columbia, Vancouver, BC  Canada V6T 1Z1}
\altaffiltext{7}{National Research Council (NRC) Fellow}
\altaffiltext{8}{Depts. of Astrophysics and Physics, EFI and CfCP, 
            University of Chicago, Chicago, IL 60637}
\altaffiltext{9}{Dept. of Physics, Brown University, 
            Providence, RI 02912}
\altaffiltext{10}{UCLA Astronomy, PO Box 951562, Los Angeles, CA 90095-1562}
\email{page@princeton.edu}

\begin{abstract}

Knowledge of the beam profiles is of critical 
importance for interpreting data from cosmic microwave background 
experiments. In this paper,
we present the characterization of the in-flight optical response of the 
{\WMAP} satellite. The main beam intensities have been 
mapped to $\le-30~$dB of their 
peak values by observing Jupiter with the satellite
in the same observing mode as for CMB observations. The beam patterns 
closely follow the pre-launch expectations. The full width at 
half maximum is a function of frequency and ranges from $0\fdg82$ at 23 GHz to
$0\fdg21$ at 94~GHz; however, the beams are not Gaussian.
We present: (a) the beam patterns for all ten 
differential radiometers and show that
the patterns are substantially independent of polarization 
in all but the 23~GHz channel; (b) the 
effective symmetrized beam patterns that result from \WMAP's compound 
spin observing pattern; (c) the effective window functions for
all radiometers and the formalism for propagating the window 
function uncertainty; and (d) the conversion factor from point source flux to
antenna temperature. A summary of the systematic uncertainties, which 
currently dominate our knowledge of the beams, is also presented. 
The constancy
of Jupiter's temperature within a frequency band is an essential check
of the optical system. The tests enable us to report a calibration of Jupiter
to 1-3\% accuracy relative to the CMB dipole. 

\end{abstract}


\section{Introduction}
The primary goal of the {\WMAP} satellite \citep{bennett/etal:2003}, 
now in orbit, is to make high-fidelity polarization-sensitive maps 
of the full sky in five frequency bands between 20 and 100 GHz. 
With these maps we characterize the properties of the cosmic microwave 
background (CMB) anisotropy and Galactic and extragalactic emission 
on angular scales ranging from the effective beam size, $\approx 0\fdg21$, 
to the full sky \citep{bennett/etal:2003b}.
{\WMAP} comprises ten dual-polarization differential 
microwave radiometers \citep{jarosik/etal:2003} fed by two 
back-to-back shaped offset Gregorian telescopes \citep{page/etal:2003}. 

Knowledge of the beam profiles is of critical importance for interpreting
data from CMB experiments. In algorithms
for recovering the CMB angular power spectrum from a map,
the output angular power spectrum is divided by the window 
function to reveal the intrinsic 
angular power spectrum of the sky. Thus, the main beam and 
its transform (or transfer function) 
directly affect cosmological analyses. Typically, the beam
must be mapped to less than $-30$~dB of the peak to achieve 1\% accuracy
on the angular power spectrum.

The {\WMAP} calibration is done entirely with the CMB dipole,
which fills the main lobes and sidelobes. Consequently, the angular 
spectrum is referenced to a multipole moment of $\ell=1$. 
The beam profile, discussed here, 
and electronic 
transfer function \citep{jarosik/etal:2003b} determine the ratio
of the window function at high $\ell$ to that at $\ell=1$.
For most other CMB experiments, 
insufficient knowledge of the beams affects both the calibration
and window function as discussed, for example, in \cite{miller/etal:2002}. 

Although it is traditional, and often acceptable, to parametrize 
beams with a single one or two-dimensional Gaussian form, such an 
approximation is not useful for \WMAP. This is because at 
the level to which the beams must be characterized, 
they are intrinsically non-Gaussian. The {\WMAP} beams 
can, however, be treated as azimuthally symmetric because each 
pixel is observed with
multiple orientations of the spacecraft. The symmetric beam approximation
is made for the first data release, avoiding many of 
the complications associated with asymmetric beams 
\citep{cheng/etal:1994,netterfield/etal:1997,
wu/etal:2001,souradeep/ratra:2001}. 

In the following we discuss how the beams are parametrized, 
how the window functions are computed, and how the uncertainties in the 
window functions are propagated through to the CMB angular power spectrum. 
The notation is summarized in Table~\ref{tab:notation}.
The sidelobe response is 
discussed in a companion paper \citep{barnes/etal:2003}. 

\section{Determination of the Beam Profiles. }

The beam profiles are determined from observations of Jupiter
while the observatory is in its nominal CMB observation mode. Jupiter 
is observed during two approximately 45 day intervals each year.
The first period\footnote{We use the JPL planetary ephemerides
available from \url{http://ssd.jpl.nasa.gov/eph\_info.html}} occurred
from 2001 October 2 until 2001 November 24 when Jupiter was at $l=193\fdg8$, 
$b=12\fdg5$, and a second period from 2002 February 8 to 2002 April 2 
with Jupiter at $l=189\fdg6$, $b=5\fdg9$, close to the Galactic 
plane and SNR IC443.  The data are analyzed 
in the same manner as the CMB data in terms of pre-whitening, 
offset subtraction, and calibration \citep{hinshaw/etal:2003}. 
The Jupiter observations are excluded from the CMB sky maps. In turn,
the CMB sky maps are used to remove the background sky signal underlying
the Jupiter maps. 

Because the distance to Jupiter changes substantially over the observing
period, all beam maps are referenced to a distance of $d_J= 5.2~$AU
and a solid angle of 
$\Omega^{ref}_J= 2.481\times10^{-8}~$sr \citep{griffin/etal:1986}. 
Jupiter is effectively a point source with temperature
$T_J = \int_\Omega T^{m}_J d\Omega/\Omega^{ref}_J
= T^{m}_J\Omega_{B}/\Omega^{ref}_J$, where $T^{m}_J$ is the amplitude
observed by {\WMAP} and $\Omega_{B}$ is the measured main beam solid angle.
A map of Jupiter is expressed as 
\begin{equation}
T({\bf n})= T_J\Omega_J^{ref}B({\bf n})
\end{equation} 
where {\bf n} is a unit direction vector and $B$ is the main beam
pattern described below. The maximum of $T({\bf n})$ corresponds
to $T^{m}_J$.
The intrinsic short term variability in Jupiter's flux is expected to be 
less than $0.02~$Jy
at 20~GHz assuming that the fluctuations scale with the nonthermal 
radio emission \citep{bolton/etal:2002}.
The measured flux is $\approx 60~$Jy in K band (23~GHz) leading to an expected
variation of $<0.1$\%. Our measurements limit any variability to 
less than 2\%.

\begin{figure*}[tb]
\plotone{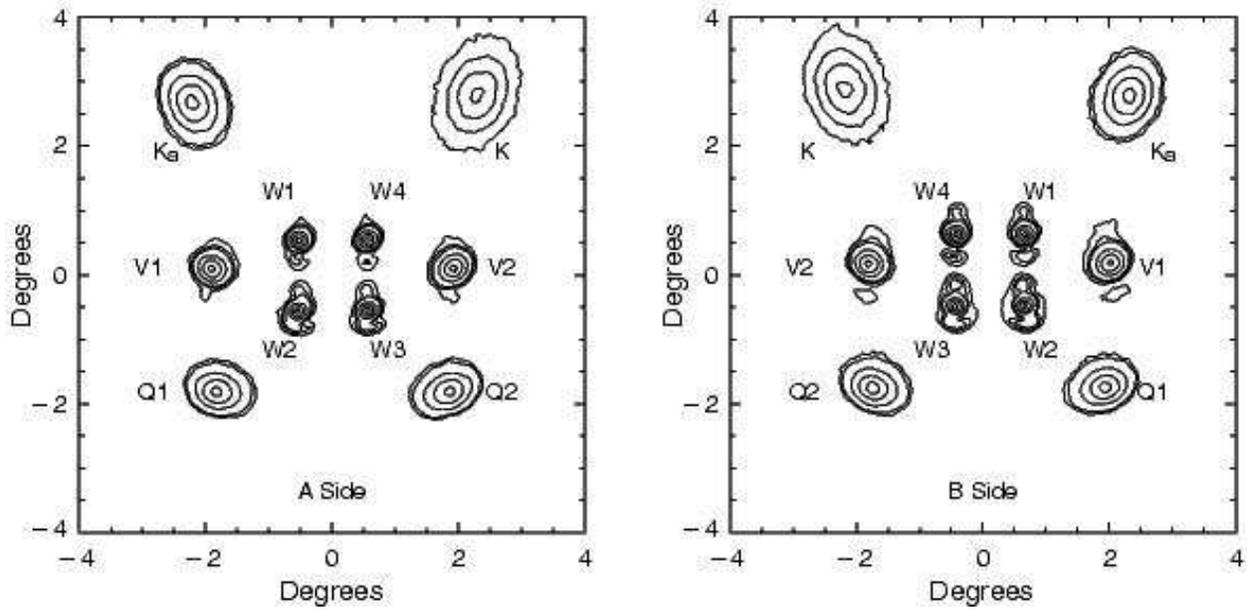}
\caption{Jupiter maps of the A and B side 
focal planes \citep{bennett/etal:2003} in the reference frame
of the observatory. The contour levels are at 0.9, 0.6, 0.3, 
0.09, 0.06, 0.03 of the peak value. W1 and W4 are the ``upper''
W-band radiometers. In W band, the lobes at the 0.09 contour level 
($\approx -10~$dB) and lower are due to surface deformations.}
\label{fig:focalplane}
\end{figure*}

Figure~\ref{fig:focalplane} 
shows the ``raw'' beams from both telescopes. A number of features
are immediately evident. As expected, all beams are 
asymmetric and the V and W band 
beams have significant substructure at the $-10$ to $-20$~dB level. 
The asymmetry results 
because the feeds \citep{barnes/etal:2002} are far 
from the primary focus. The substructure 
arises because the primary mirror distorts upon cooling with an 
rms deviation of $\sigma_z=0.024~$cm and correlation length of 
$l_c\approx 10~$cm \citep{page/etal:2003}.   

The Jupiter data are analyzed both as maps binned with 
$2.4^\prime\times 2.4^\prime$ pixels and as time ordered data (TOD).
These data products are analyzed separately.
In addition, full flight simulations are used to 
test the software.

Figure~\ref{fig:beams} shows the beams in profile after symmetrization. 
In the Jupiter map analysis, the symmetrization procedure consists of 
smoothly interpolating the beam to $0.015^\prime\times0.015^\prime$ pixels 
with a 2-D spline and then azimuthally averaging
in rings of width $1.2^\prime$. 
Due to noise, the maximum value in a map is often
not on the best symmetry axis, though it is generally within one pixel of it.
The symmetrized beam has the same solid angle as the
raw beam to within $0.3$\%. The normalized symmetrized beam is called $b^S$.  
In the TOD analysis, the data are fit to a series
of Hermite functions (\S{\ref{sec:winfunc}}) according 
to their angular separation from a predetermined centroid. The 
best centroid is determined iteratively. The analysis 
includes the effects of the sampling integration and pre-whitening procedure
\citep{hinshaw/etal:2003} but does not have
the intermediate step of mapmaking and thus is independent of the
$2.4^{\prime}$ pixelization. 
There are low signal to noise modes in the Jupiter maps that do not affect
the window functions and to which the Hermite 
method is insensitive. Generally,
the Hermite beams are used for $\ell$-space quantities, such 
as the window function, and the Jupiter maps are used for real
space calculations.  

\subsection{Beam solid angles and uncertainties\label{sec:bu}}

Linking the calibration of observations of the CMB at $\ell=1$ 
to compact sources
at $\ell\approx1000$ requires knowledge of the beam profile over
a large range of angular scales. The width in $\ell$-space can be 
characterized by the total beam solid angle, 
$\Omega_A =\int b_T\,d\,\Omega$,
where $b_T$ is the full sky beam profile normalized to unity 
on the boresight. In our analysis, the beam is divided into two parts: 
$\Omega_A =\Omega_B+\Omega_S$
where $\Omega_B$ is the main beam solid angle and $\Omega_S$ is the 
portion associated with the sidelobes. Ideally $\Omega_S=0$ but this
is not an appropriate approximation for {\WMAP}, especially in K band.
Generally, $\Omega_B$ is found by normalizing a map to the 
best fit peak value, and then 
integrating out to some radius. For {\WMAP}, $\Omega_B$ does not 
reach a stable value as the integration radius is increased
because of the imperfect knowledge of the background. This may be seen 
with the following estimate. The net statistical uncertainty in the
{\WMAP} data in a patch 
of radius $3\deg$ is 
$\approx6~\mu$K (the noise is $\approx0.85$~mK in 
each $2.4^\prime\times 2.4^\prime$ pixel). 
If the background could be removed to this level, the resulting solid angle
temperature product would be $T\Omega=5\times 10^{-2}~\mu$K\,sr. 
By comparison, $T_J^m\Omega_B\approx 4~\mu$K\,sr in W-band
($T_J^{m}\approx 200~$mK and $\Omega_B\approx 2\times 10^{-5}~$sr ). 
Given that both
the region near Jupiter and the region being compared to must be known to 
this level and that $T\Omega$ grows with radius,
deviations of a few percent are not unexpected. As the signal to noise
improves throughout the mission, this limitation will be alleviated. 

In order to derive a consistent set of solid angles,
we define a cutoff radius, $\theta_{Rc}$, out to which
the solid angle integration is performed. As shown in 
\cite{page/etal:2003}, the global properties of the beam may be 
{\it modeled} 
by a core plus a ``Ruze pattern'' \citep{ruze:1966} that is a function of the 
surface correlation length and rms roughness. We assume that $l_c$
is unchanged from the pre-launch measurements at 70~K.
(The in-flight temperatures of the primaries are 73~K and 68~K for the 
central and top thermometers for both A and B sides.) Table 
\ref{tab:gain_budget} shows how the main beam forward gain, $4\pi/\Omega_B$,
is reduced by the sampling and surface deformations. The agreement between the
measurements and the expectations is evidence that the main beam 
effects have been accounted for and that the
in-flight $\sigma_z$ is consistent with the ground-based measurement.
Physical models of the optics, in \S{\ref{sec:pmodels}}, 
give further evidence for this.

\begin{figure*}[tb]
\plotone{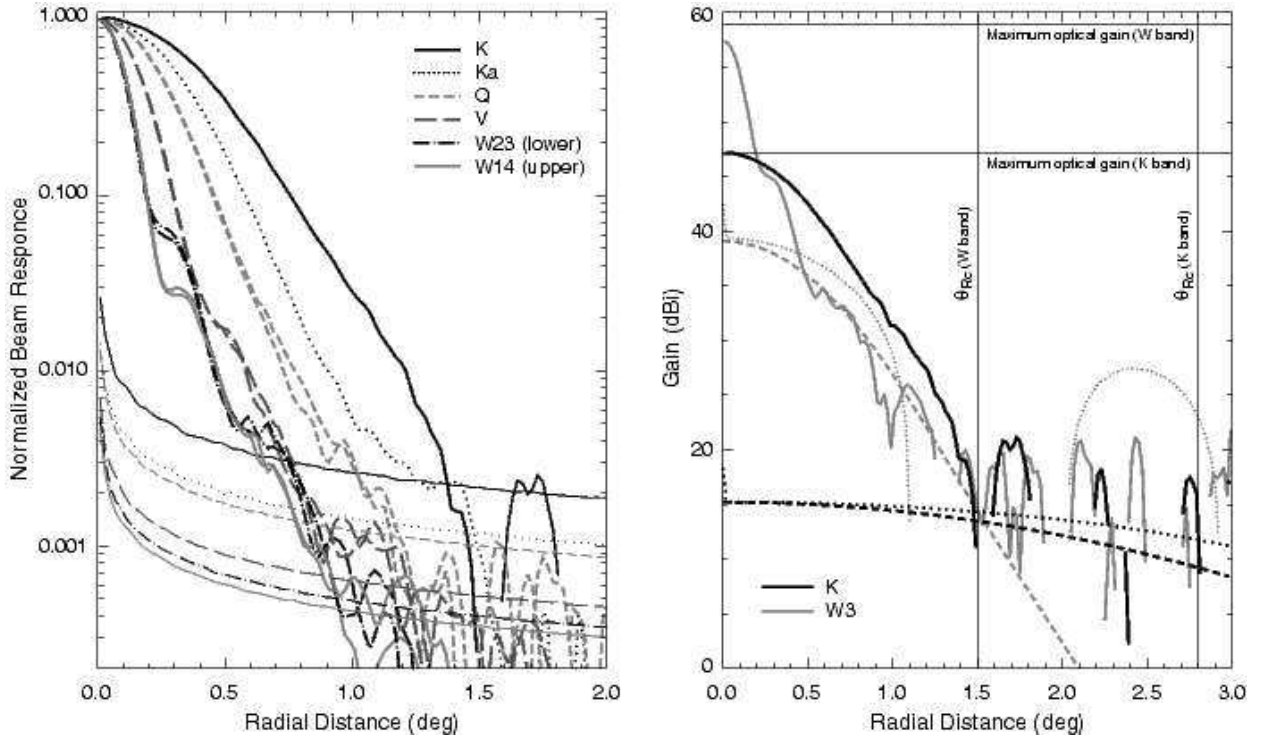}
\caption{Left: The symmetrized beams (normalized at unity) and noise levels 
(below) from 
two seasons of Jupiter observations. Both polarizations have been combined.
The noise rises at
small radii because there are fewer pixels over which to average.
With four years of observations, the noise level will be reduced by a 
factor of two. Right: 
The K (black) and W3 (grey) symmetrized beam profiles
with their associated Ruze patterns (\S{\ref{sec:bu}}).
The noise level is at 20 dBi in all bands as seen in the plot
(missing data corresponds to negative values). The maximum optical gains
are 47.1 and 59.3~dBi in K and W bands as indicated by the horizontal 
lines. Table~1 shows the gain budget.
The dashed lines are the Ruze patterns assuming a Gaussian shaped
distortion with the parameters in the text. The lighter shaded dotted lines
that meet the dashed lines at $\theta=0$ are
for a tophat shaped distortion. In W band, the tophat prediction, 
which has a prominent lobe at $\theta=2\fdg5$ clearly does not fit the data.
Plots for W14 show the Ruze pattern to be above the beam profile for
$\theta<1^{\circ}$ suggesting the magnitude of the deformations
is not greater than
those we use. However, some fraction of $\Omega_B$ could be at or near 
the noise level for $1\fdg5<\theta<2\fdg0$.
The vertical straight lines indicate 
the cutoff radii, $\theta_{Rc}$, for the Gaussian distortion model. } 
\label{fig:beams}
\end{figure*}

We define $\theta_{Rc}$ as the radius that contains 
at least 99.8\% of the {\it modeled} main beam solid angle. 
These values are
2\fdg8, 2\fdg5, 2\fdg2, 1\fdg8, 1\fdg5 respectively for K through W bands
and agree with values derived from the physical model of the optics. 
Figure~\ref{fig:beams} shows examples of the K- and W-band 
beams and their Ruze patterns.
The solid angles resulting from the integrals over the Jupiter maps
with a cutoff radius of  $\theta_{Rc}$ are given in Table~\ref{tab:fpa_beams}.

The uncertainty in the main beam solid angle primarily affects the
determination of the flux from point sources and spatially 
localized features in real space. Solid angle uncertainties
affect the angular spectrum through
the width of the $\ell$-space passband as discussed in
\S{\ref{sec:winfunc}}. The uncertainty is assessed three ways:

\begin{enumerate} 

\item The solid angles are determined from simulations 
that have been processed in the same way the flight data
are computed. The rms deviation
between the input and recovered solid angle for 20 measurements 
(10 on the A side and 10 on the B side) is 1.1\%. The formal statistical
uncertainty is between 0.7\% and 1\% depending on band.

\item The scatter in the derived temperature of Jupiter is found
for all detector/reflector combinations. Each 
feed is mapped in two polarizations by two detectors during two seasons
for a total of eight measurements. The statistical uncertainty of the 
mean of the eight measurements is 2.6\% in K band and $\approx 1.1$\%
in Ka through W bands. The K beam result has a relatively low amplitude, 
$\approx 13.5$~mK, and is wide, making it difficult to measure. It is
also the most susceptible to the effective frequency and to 
incompletely subtracted Galactic 
contamination. 
 
The systematic uncertainty is also determined from the Jupiter maps.
The uncertainty on Jupiter's measured amplitude 
is $\approx 0.5$\% and is always subdominant to the
uncertainty in the solid angle. We assume that Jupiter's temperature
is the same for all measurements within a band. (The intraband effective
frequencies are close enough to be insensitive to Jupiter's spectrum.)
The uncertainty of $\Omega_{B}$ is increased until $\chi^2/\nu=1$
for fits to Jupiter's temperature within a band. 
This results in uncertainties of 2.6\%, 1.2\%, 1.2\%, 1.1\%, \& 2.1\% 
per DA\footnote{A differencing assembly (DA) \citep{jarosik/etal:2003}
comprises two polarization sensitive radiometers. There 
are 1, 1, 2, 2, \& 4 DAs in K through W bands repectively.} per side for K
through W bands.

\item The solid angles are recomputed by direct integration
after increasing $\theta_{Rc}$ 
to 3\fdg7, 3\fdg3, 2\fdg9, 2\fdg4, 2\fdg0 respectively for K through W bands.
For the 10 DAs on both A and B sides, the rms deviation between
the original $\theta<\theta_{R_c}$ and recomputed 
solid angles is 0.8\% with no clear 
trend in the sign of the deviation except in W-band. In W-band,
the solid angles with the increased $\theta_{Rc}$ are systematically
larger by 0.8\% on average, suggesting a potential bias. 
The most likely explanation is that the shape of the surface
distortions is not Gaussian as assumed for the Ruze model.
From Figure~\ref{fig:beams}, one sees
that the level of the potential contribution is at $\approx -35$~dB,
just beyond $\theta_{Rc}$. We term this region the beam pedestal. 
To account for the bias,
the W-band solid angles are increased by 0.8\% over the nominally
computed value and assigned an additional uncertainty of 0.4\%,
added in quadrature. The net uncertainty is still 2.1\%. 
This increase is accounted for in Table 1 and discussed
in the context of the window functions in \S{\ref{sec:wluncert}}.
For the year-one data release, we treat this as a systematic deviation from 
our model and account for it in the analysis. 
Future analyses, with more data, will treat this effect with
a more comprehensive beam model (\S{\ref{sec:pmodels}}) and 
will determine the source of the pedestal in the beam.
 
\end{enumerate}

The uncertainties in the solid angles used throughout the analysis
encompass the systematic effects in items 2 and 3. 
These uncertainties should be interpreted as ``1$\sigma$.'' 

\subsection{Sidelobes}

The distinction between the main beams and the sidelobes 
is at some level an arbitrary definition.
The structure that holds the feed horns scatters radiation
into a large region around the main beams. In other words,
the main beam does not contain the total solid angle of the full sky 
beam (Table 7, \cite{page/etal:2003}). The fraction of
the total solid angle outside $\theta_{Rc}$ is 0.037, 0.012, 0.012,
0.0022, and 0.001-0.003 in K through W bands respectively.
For example, in K band, the region with $\theta<\theta_{Rc}$ 
contains 99.8\% of 
the modeled main beam solid angle ($\Omega_B$) 
but only 96.4\% of the total solid angle ($\Omega_A$).  

The sidelobes have two effects on the interpretation of the data.
The first arises from Galactic contamination. 
As shown in \cite{barnes/etal:2003}, the sidelobe leakage 
affects primarily $\ell<20$ and is not significant for Ka, Q, V, and W bands.
The second arises from the sidelobe contribution to the calibration
of features at $\ell>20$. For example, point sources are 
detected only in the main beam 
and the measured temperature profile of a point source corresponds to only 
a main beam calibration. The dipole, however, is a full beam calibrator.
Thus, to obtain the true flux of a point source, to a good approximation 
one multiplies the flux as calibrated by the dipole by  1.037, 1.012, 1.012,
1.002, and 1.003 in K through W bands respectively. 

For the year-one release, only the K-band map is corrected
for the Galactic sidelobe contribution. However, all the point source
fluxes in \cite{bennett/etal:2003c} and Jupiter fluxes given below
have been corrected in all bands. The uncertainty of the correction
is taken as half the correction factor, or 2\%, 0.5\%, 0.5\%, 
0.1\%, and 0.2\% in K through W bands respectively. 
 
\subsection{Effective Frequencies}

Because of \WMAP's broad frequency bandwidth, sources with
different spectra have different effective frequencies.
The effective frequency for a source that is small compared
to the beam width \citep{page/etal:2003} is:
\begin{equation}
\nu_e \equiv  
{ \int \nu f(\nu) G_{m}(\nu)\nu^{-2}\sigma(\nu) d\nu \over
\int f(\nu)G_{m}(\nu)\nu^{-2}\sigma(\nu) d\nu }
\label{eq:nubb}
\end{equation}
where $f(\nu)$ is the measured radio frequency (RF) passband 
\footnote{\cite{jarosik/etal:2003} uses $r(\nu )$ where this paper
and \cite{page/etal:2003} use $f(\nu )$.} and $G_{m}(\nu)$
is the maximum (forward) gain. The spectral dependence
of the source is $\sigma(\nu)\propto \nu^{\alpha}$
with $\alpha$ the spectral index ( $\alpha\approx -0.7$ for synchrotron, 
$\alpha \approx -0.1$ for free-free,
$\alpha=2$ for Rayleigh-Jeans, and $\alpha\approx 4$ for dust) or 
$\sigma(\nu)\propto \nu^4\exp(h\nu/k_BT_{CMB})/(\exp(h\nu/k_BT_{CMB})-1)^2$ 
for the CMB. (The variation of loss across the band is negligible.)
For full beam sources, such as the CMB or the calibrators used in ground
testing, the small dependence on the forward gain should not be included
and the central frequencies in \cite{jarosik/etal:2003} should be used.  
At high $\ell$ or for point sources,
the effective center frequency, and therefore the thermodynamic to 
Rayleigh-Jeans correction, changes. The magnitude and sign of the 
change depend on the relative weights of the radiometer passbands 
and forward gain. This small effect ($<0.2~$GHz or $<0.1$\%) in 
the conversion was not included in the year-one maps. 

The passbands for the 
two polarizations in a DA
differ slightly  \citep{jarosik/etal:2003}. Thus, the two 
polarizations for one telescope (e.g., A side)  have different passbands.
On the other hand,
the optical gain as a function of frequency is the same for
both polarizations of one telescope but differs between the A and B sides.
In Table~\ref{tab:fpa_beams}, for the CMB, we give the effective 
frequencies 
from equation~\ref{eq:nubb} for the average of the two RF passbands 
for the A and B sides separately. We also give the effective frequencies
for both polarizations separately (Table 11, \cite{jarosik/etal:2003} )
for a source that fills the beam. The table shows that the effect of 
the optics on the passbands is small.

\subsection{Temperature of Jupiter}

The observations of Jupiter and the CMB dipole with {\WMAP} result in
a calibration of Jupiter
calibrated with respect to the CMB dipole. After coadding the 
data over polarization and season, 
a fit is made to Jupiter's temperature. Before the fit is done, 
a correction is applied. 
The loss in the input optics on the A and B sides differs by $\approx~1$\% 
(Table 3, \cite{jarosik/etal:2003}). Therefore, Jupiter and the CMB
do not have the same apparent temperature when measured through the A
and B telescopes. The signature of the effect is that the 
average of the A-side temperature data is offset from that of 
the B-side data. This effect is accounted for in the year-one sky maps
but not in the Jupiter maps.
After the A/B imbalance, sidelobe corrections, and the
W-band pedestal correction, all measurements of
$T_J$ in one microwave band are fit to a single temperature.
We find that $T_J$, in brightness temperature, 
is given by $134\pm4$, $146.6\pm2.0$, $154.7\pm1.7$, 
$163.8\pm1.5$, $171.8\pm1.7~$K in K through W bands respectively. 
The uncertainty is dominated by 
the uncertainty in the solid angles, in the sidelobe corrections,
and in the 0.5\% intrinsic calibration uncertainty \citep{hinshaw/etal:2003}.
These values are the temperature one would measure by comparing the flux
from Jupiter to that from blank sky. To obtain the absolute
brightness temperature, one must add to these the brightness temperature
of the CMB (2.2~K, 2.0~K, 1.9~K, 1.5~K, and 1.1~K
in K through W bands, respectively). A future paper will 
compare this result to other measurements and assess the stability and
polarization characteristics more completely. 

\subsection{Physical Model of Beams}
\label{sec:pmodels}

The dominant surface deformation that leads to the distortions 
of the main beams comes from the ``H-shaped'' backing structure
that holds the primary mirrors. A simple Fourier transform 
of the aperture with an additive H-shaped distortion 
reproduces many of the features in Figure~\ref{fig:focalplane},
but photogramatic pictures \citep{page/etal:2003} of the cold surface
indicate that the surface structure is more complicated. 
For the purposes of the year-one analysis, 
the Gaussian distortion model is sufficient,
but for future analyses a more accurate model is desired.

The full beam is modeled
using a physical optics code (DADRA, \cite{YRS:DADRA}) that predicts the 
beam profile given the detailed physical shape of the optics.
The primary surface deformation is parametrized with a set of
Fourier modes the amplitudes of which are the fit coefficients. 
(In the pre-launch cryogenic tests, there was 
no evidence for a significant change in shape of the secondary.) 
A minimization loop finds the surface shape parameters that
simultaneously fit the two V-band and four W-band beam maps
of Jupiter. 

\begin{figure*}[tb]
\plotone{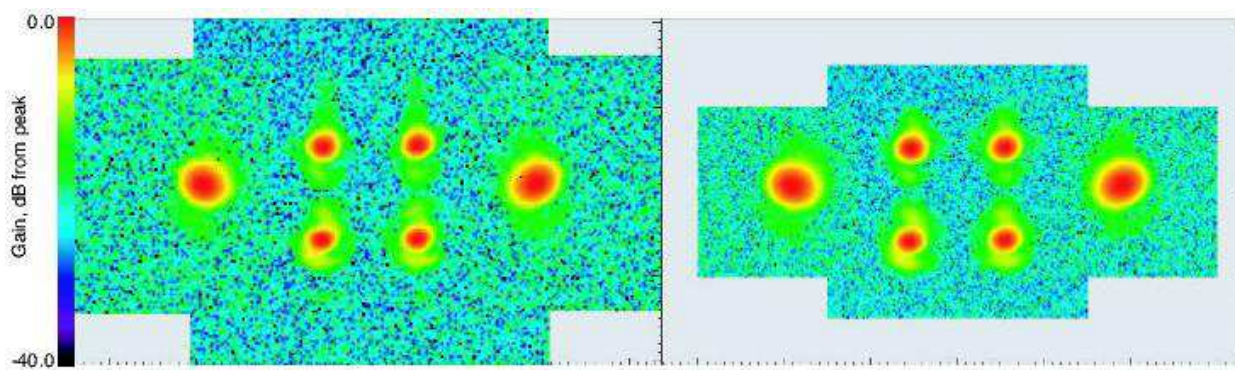}
\caption{Left: A mosaic of the A-side W and V-band measured beams.
One should focus on the main beams areas. Different noise levels 
in the constituent mosaics lead to apparent artifacts away from the 
beam centers.
Right: Model of the A-side beams based on the physical optics calculations
described
in the text. The same surface is used for all beams. 
Most of the features in the measured beams are reproduced
in the model indicating that the source of the distortions has
been identified. The noise from the measurement has been added to 
model on the right to make the comparison more direct.
The separation between different W-band beams is $1\fdg1$, less than the 
cutoff radius for the determining the W-band solid angles.}
\label{fig:pred_beams}
\end{figure*}

The program is computationally intensive because the 
full physical optics calculation for all beams is recomputed 
for each iteration.  A qualitative comparison of the measured 
and modeled beam patterns is shown in Figure~\ref{fig:pred_beams} 
from which it is clear that the amplitude and phase of much of 
the surface deformation has been identified. 
The program has not run long enough to converge in all bands, 
so it is not yet used in the beam analysis. At this stage, though,
the model gives us confidence that we are not missing significant
components of $\Omega_B$ and that our interpretation of the beams
is correct.

\subsection{Temperature Stability and Reflector Emissivity}

The {\WMAP} orbit in the L$_2$ environment results in extremely
good thermal stability. The instrument has an array of thermometers 
on the optical components with sub-millikelvin 
resolution \citep{jarosik/etal:2003b}.
By binning the temperature data synchronously with the spin rate and
the position of the Sun, we detect a synchronous 
thermal variation in the optics. The top and middle of the primary
have a peak-to-peak amplitude of $\approx 0.23$~mK. We believe
this is due to scattering of
solar radiation off the rim of the solar array.
The tips of the secondary mirrors show a maximal
variation of 0.04~mK peak-to-peak, with the rest of the secondaries
much less.
There is no evidence for any thermal variation due to illumination by 
the Earth or Moon. The 0.23~mK is well below the conservative bound
of 1.5~mK in \cite{page/etal:2003} and below the 0.5~mK rms requirement
in the systematic error budget. 

We bound the emissivity, $\epsilon$, of the reflectors using a similar method.
After subtracting the CMB dipole from the TOD, the 
radiometric signal is coadded in Sun-synchronous coordinates
following the method outlined in \cite{jarosik/etal:2003b}.
The {\it net} spin synchronous radiometric signal 
detected is 0.4~$\mu$K peak-to-peak (0.014~$\mu$K rms) in the combined
W and V bands. 
Therefore, an upper bound on the emissivity of the surface is 0.002.
The predicted emissivity is $0.0005$.

\subsection{Polarization from Optics}

Each DA measures two orthogonal differential polarizations
from each pair of feeds (e.g., K11A and K11B form one difference and
K12A and K12B form the other difference). The Stokes Q and U components
are found by differencing these two signals in the time stream, determining
the components relative to a fixed direction on the sky,
and then producing a map with the mapmaking algorithm 
\citep{kogut/etal:2003, hinshaw/etal:2003}.
{\WMAP} was designed to have cross-polar leakage of $<-22$~dB in all
bands \citep{page/etal:2003} to enable a measurement of the polarization
of the CMB. This specification was met and demonstrated in pre-launch
ground tests with a polarized source. 
 
To assess the degree of the similarity of the two polarization channels, 
we take the A-side beam maps for both polarizations from 
one feed (e.g., KA11 ('P1')
and KA12 ('P2')), difference them, integrate
over the difference map, and then divide by $T\Omega$ from Jupiter. 
The resulting fractional signal for the A side is 
8.1\%, 2.5\%, 0.2\%, 0.4\%, -2.8\%, -2.7\%, 0.3\%, 0.5\%,
1.6\%, and 0.9\% in K1 through W4 bands.
For the B side it is 6.5\%, -0.1\%, -3.3\%, 4.0\%, 0.4\%, 1.0\%, -0.5\%,
-0.1\%, 2.7\%, -0.5\%.
The statistical uncertainty for the difference in polarizations is 
$\approx 1.2$\% and the measured rms of all Ka through W values is 1.8\%.
For a polarized source, the sign of the difference changes when P1-P2
is determined on the A and B sides.

The difference between the beams (P1-P2) from the Jupiter maps 
is larger than expected but there
are no clear trends in Ka through W bands and no clear detections 
of polarization.
For K band, there is a clear excess at a level larger than can be 
attributed to the optics. As Jupiter is nearly a thermal source, 
the difference is not due to a frequency mismatch \citep{kogut/etal:2003}
between the dipole and Jupiter. However, the difference in effective
frequencies
of 1.0~GHz (P1 and P2, Table~\ref{tab:fpa_beams}) 
leads to a difference in solid angle of order 10\% 
(Table 5, \cite{page/etal:2003}), enough to explain the effect.
The net effect is smaller in the higher frequency bands because 
of the low edge taper.  For the year-one analysis, the slight 
difference in the effective frequency
for the bands is not corrected. Instead, the uncertainty
is absorbed in the uncertainty of the average solid angle.

The CMB is polarized at the $\approx 5$\% level, in temperature units,
and the temperature-polarization correlation is at the $\approx 15$\% level.
The CMB polarization signal comes from scales larger than the size of the beam
and so no particular band mismatch affects the CMB polarization results.
In addition, the CMB signal is derived from the sum 
over multiple bands.  

\section{Calculation of window functions}
\label{sec:winfunc}
The characteristics of the CMB are most frequently expressed as an
angular spectrum of the form $l(l+1)C_l/2\pi$ \citep{bond:1996}
where $C_l$ is the angular power spectrum of the temperature:
\begin{equation}
T({\bf n}) = \sum_{\ell,m}a_{\ell m}Y_{\ell m}({\bf n}),
~~~  \langle a^{*}_{\ell' m'}a_{\ell m}\rangle
=\delta_{\ell' \ell}\delta_{m'm}C_\ell
\end{equation}
where {\bf n} is a unit vector on the sphere and $Y_{\ell m}$ is a 
spherical harmonic.

The beam acts as a spatial low-pass filter on the angular
variations in $T({\bf n})$ such that the variance of a noiseless
set of temperature measurements is given by 
\begin{equation}
C(0)\approx\sum_{\ell}{(2\ell+1)C_\ell \over 4\pi } w_\ell
\end{equation}
where $w_\ell$ is the window function which encodes the beam smoothing.
It is normalized to unity at $\ell=0$ as discussed below. 

The window functions are expressed following the conventions
of \cite{white/srednicki:1995}. The window function depends on the
mapping function $M({\bf n},{\bf n}')$ which describes how the
experiment convolves the true sky temperature $T({\bf n}')$ into the observed
temperature ${\widetilde T}({\bf n})$,
\begin{equation}
{\widetilde T}({\bf n}) = \int d\Omega_{{\bf n}'}M({\bf n},{\bf n}')T({\bf n}').
\end{equation}

For \WMAP, $M({\bf n},{\bf n}')$ for one feed is a weighted average of
the beam response $B({\bf n},{\bf n}')$ that accounts for the
smearing due to the finite arc scanned over an integration period and the
azimuthal coverage of the observations in each pixel. The symmetrized beam,
$B^S$, is an excellent approximation to the mapping function.

The full window function is given by:
\begin{eqnarray}
w_\ell({\bf n}_1,{\bf n}_2)
 &\equiv&  \int d\Omega_{{\bf n}'_1} \int d\Omega_{{\bf n}'_2}
         M({\bf n}_1,{\bf n}'_1)M({\bf n}_2,{\bf n}'_2)
         P_\ell({\bf n}_1'\cdot{\bf n}_2')\nonumber\\
 &= &{4\pi \over (2\ell+1)}\sum_{m=-\ell}^{+\ell}
    \int d\Omega_{{\bf n}'_1}  \int d\Omega_{{\bf n}'_2}
   \, M({\bf n}_1,{\bf n}'_1)\times \nonumber\\ 
   &  &\, M({\bf n}_2,{\bf n}'_2)
   \, Y^{*}_{\ell m}({\bf n}_1')\,Y_{\ell m}({\bf n}_2')
\end{eqnarray}
where $P_\ell$ is a Legendre polynomial.
The primary quantity of interest is the window function at zero lag, 
$w_\ell({\bf n},{\bf n})$. In this case the total variance of the data,
$C(0)$, is the sum of the power in each spherical harmonic 
weighted by the window. This is directly analogous to 
low-pass filtering. If the
mapping function is independent of celestial position, then 
 
\begin{eqnarray}
w_\ell & = & {4\pi \over (2\ell+1)}\sum_{m=-\ell}^{+\ell}
             \left| \int d\Omega_{{\bf n}} 
             \, M({\bf n},{\bf n})\, Y^{*}_{\ell m}({\bf n})\right|^2 \\
\label{eq:w_yy}
       & = & {4\pi \over (2\ell+1)}\sum_{m=-\ell}^{+\ell}
             \left| m_{\ell m} \right|^2
\end{eqnarray}
where the $m_{\ell m}$ are the harmonic coefficients of the mapping function.

If the beam is azimuthally symmetric, a further simplification can be made:
$\sum_{m=-\ell}^{+\ell} \left| m_{\ell m} \right|^2=(2\ell+1)b_l^2/4\pi$
where $b_\ell$ is given by the Legendre transform of the beam
\begin{equation}
b_\ell= 2\pi\int b^S(\theta )
P_\ell(\cos\theta)d(\cos\theta)/\Omega_B.
\label{eq:bl}
\end{equation}
Thus $w_\ell=b_\ell^2$. 
For symmetric beams, $\theta$ is used instead of 
${\rm acos}({\bf n}\cdot{\bf n'})$.
For a symmetric Gaussian of width $\sigma_b$, 
we find
\begin{eqnarray}
B_\ell &\approx& 2\pi\sigma_b^2 e^{-\ell(\ell+1)\sigma_b^2/2} \cr
&=& \Omega_B e^{-\ell(\ell+1)\sigma_b^2/2}
\end{eqnarray}
where $B_\ell$ is the Legendre transform with units of sr.
Because the instrument is
calibrated with a dipole signal, the $B_\ell$ 
should be normalized at $\ell=1$. In practice, this is indistinguishable 
from normalizing at $\ell=0$.
Thus, $b_\ell=B_\ell/B_{\ell=0}=B_\ell/\Omega_B$ and is dimensionless.

Although {\WMAP} is
intrinsically differential, there is essentially no overlap of the
beams from opposite sides, so the window functions from each beam 
may be treated
independently and then combined in the end. Thus, the window function 
for the differential signal is derived from the weighted symmetrized beam

\begin{equation}
b^S(\theta) = 
{\Omega^S\over 2}
 \biggl({ (1-\overline{x}_{im})|b^{S,A}|\over \Omega^A_B} 
     + { (1+\overline{x}_{im})|b^{S,B}|\over \Omega^B_B}\biggr)
\label{eq:bs}
\end{equation}
where $\Omega^A_B$ and $\Omega^B_B$ are the main beam solid angles
for the A and B side beams, $\Omega^S$ is the effective solid angle 
of the combined beam, and $\overline{x}_{im}$ is $\approx 0.01$ and
corrects for the A/B imbalance \citep{jarosik/etal:2003}
\footnote{The $\overline{x}_{im}$ is the average of the values for 
the two polarizations.}. We use the absolute values to indicate that
both beams are treated as positive in this equation.
As before, the superscript $S$ denotes a symmetrized beam.

\subsection{Window Functions and Their Uncertainties.}

We compute the window function for the CMB analysis 
from $b^S$ using an expansion
of the symmetrized beam in Hermite polynomials.
Hermite functions are a natural basis
for the beam as they parametrize deviations from Gaussanity.
The expansion also naturally gives the covariance matrix
for the $B_\ell$ as well. 

The Hermite expansion is given by
\begin{equation}
b^S(\theta) =  \exp (-\theta^2/2\sigma_h ^2)
\sum_{i=0}^{m_h}a_{2i}H_{2i}(\theta/\sigma_h)
\label{eq:hermite_exp}
\end{equation}
where $\theta$ is angular distance from the beam center,
$\sigma_h$ is the Gaussian width of the beam, and $H_{2i}$ is
the Hermite polynomial of order $2n$ (Chapter 22, 
\cite{abramowitz/stegun:HOMF} ).
The parameters of the expansion are given in
Table~\ref{tab:wins}. A fit is made of the TOD 
to equation~\ref{eq:hermite_exp}. From the fit, the
$m_h$ coefficients $a_{2i}$ and the $m_h\times m_h$
covariance matrix $C^{aa'}$ are found.
The expansion coefficients, $a_{2i}$, are normalized to account for
the measured temperature of Jupiter and the normalization of $H_{2i}$.

The transfer function is computed
separately for each Hermite polynomial following equation~\ref{eq:bl}:
\begin{equation}
B_{\ell i} =  
2\pi \int \exp (-\theta^2/2\sigma_h ^2)
H_{2i}(\theta/\sigma_h)P_\ell(\cos\theta )d(\cos\theta )
\end{equation}
so that the full transfer function is
\begin{equation}
B_{\ell} = \sum_{i=0}^{m_h} a_{2i}B_{\ell i} 
\label{eq:bl_hermite_exp}.
\end{equation}

\begin{figure*}[tb]
\plotone{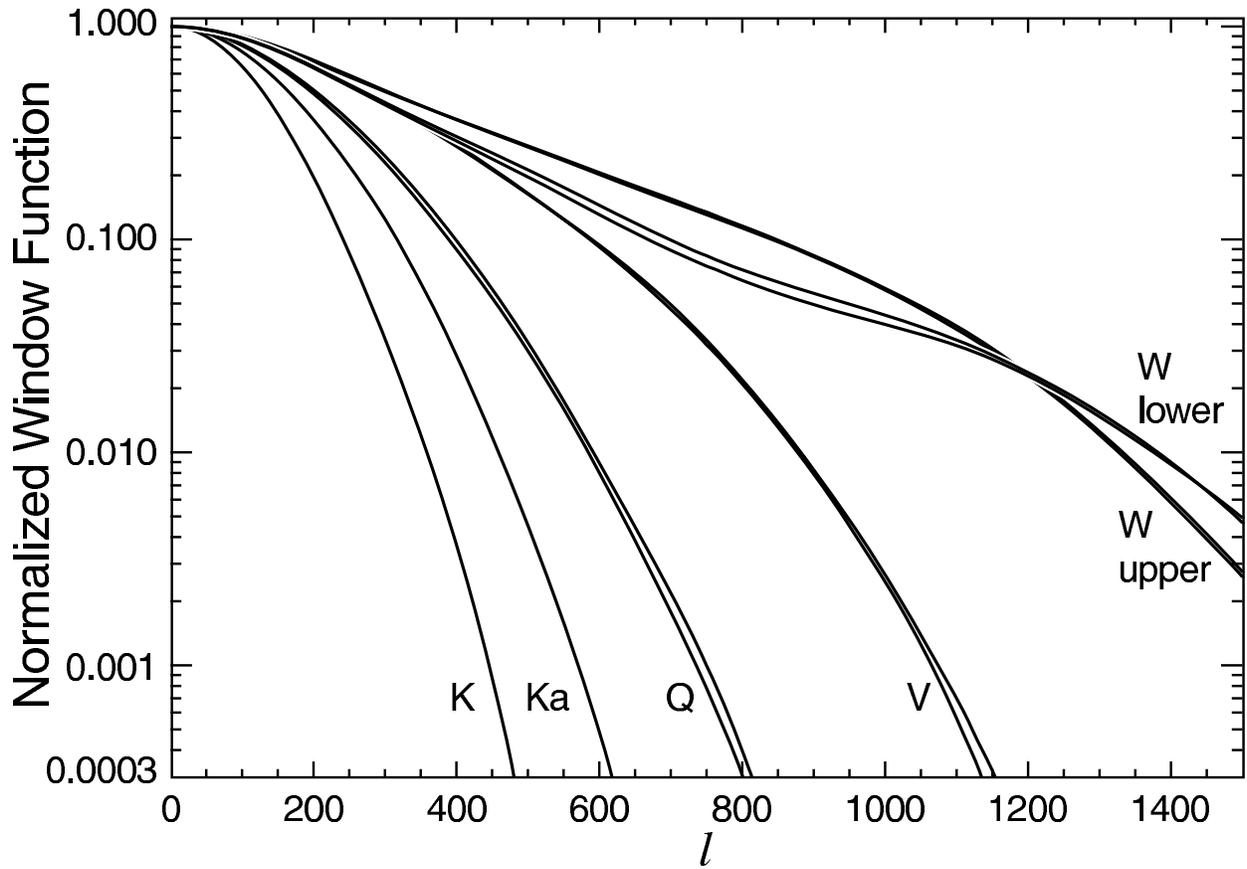}
\caption{The ten window functions, $w_\ell$, computed 
from the Hermite expansion. 
The window functions for the two polarizations in each feed are nearly
indistinguishable at the resolution of the plot.}
\label{fig:beams_wins}
\end{figure*}

The window functions based on equation~\ref{eq:bl_hermite_exp}
are shown in Figure~\ref{fig:beams_wins}.
From equation \ref{eq:bl_hermite_exp}, one can determine the
covariance matrix between $B_\ell$ and $B_\ell'$ as: 
\begin{equation}
\Sigma_{\ell \ell'}^B = 
\sum_{i,j=0}^{m_h}{\partial B_\ell\over \partial a_{2i}} C^{aa'}_{ij}
{\partial B_{\ell'}\over\partial a_{2j}}
=
\sum_{i,j=0}^{m_h}B_{\ell i}C^{aa'}_{ij}B_{\ell' j}.
\end{equation}
This matrix has units of sr$^2$, is independent of Jupiter's temperature,
and is largest in magnitude at low $\ell$.

Because of the dipole calibration, there is effectively only calibration
uncertainty at $\ell=1$.
This is accommodated in the formalism by normalizing the
beam covariance matrix at $\ell=0$, $b_\ell=B_\ell/B_0$, as 
shown in Appendix A. We find
\begin{eqnarray}
\Sigma_{\ell \ell'}^b &=& 
\sum_{i,j=0}^{m_h}{\partial b_\ell\over \partial a_{2i}} C^{aa'}_{ij}
{\partial b_{\ell'}\over\partial a_{2j}}\cr
&=&
{1\over{(\Omega^S)^2}}\left(
\Sigma_{\ell \ell'}^B            + 
b_\ell b_{\ell'}\Sigma_{00}^B   -
b_\ell\Sigma_{0\ell'}^B         - 
b_{\ell'}\Sigma_{\ell 0}^B           
\right) .
\label{eq:fll}
\end{eqnarray}

\begin{figure*}[tb]
\epsscale{0.8}
\plotone{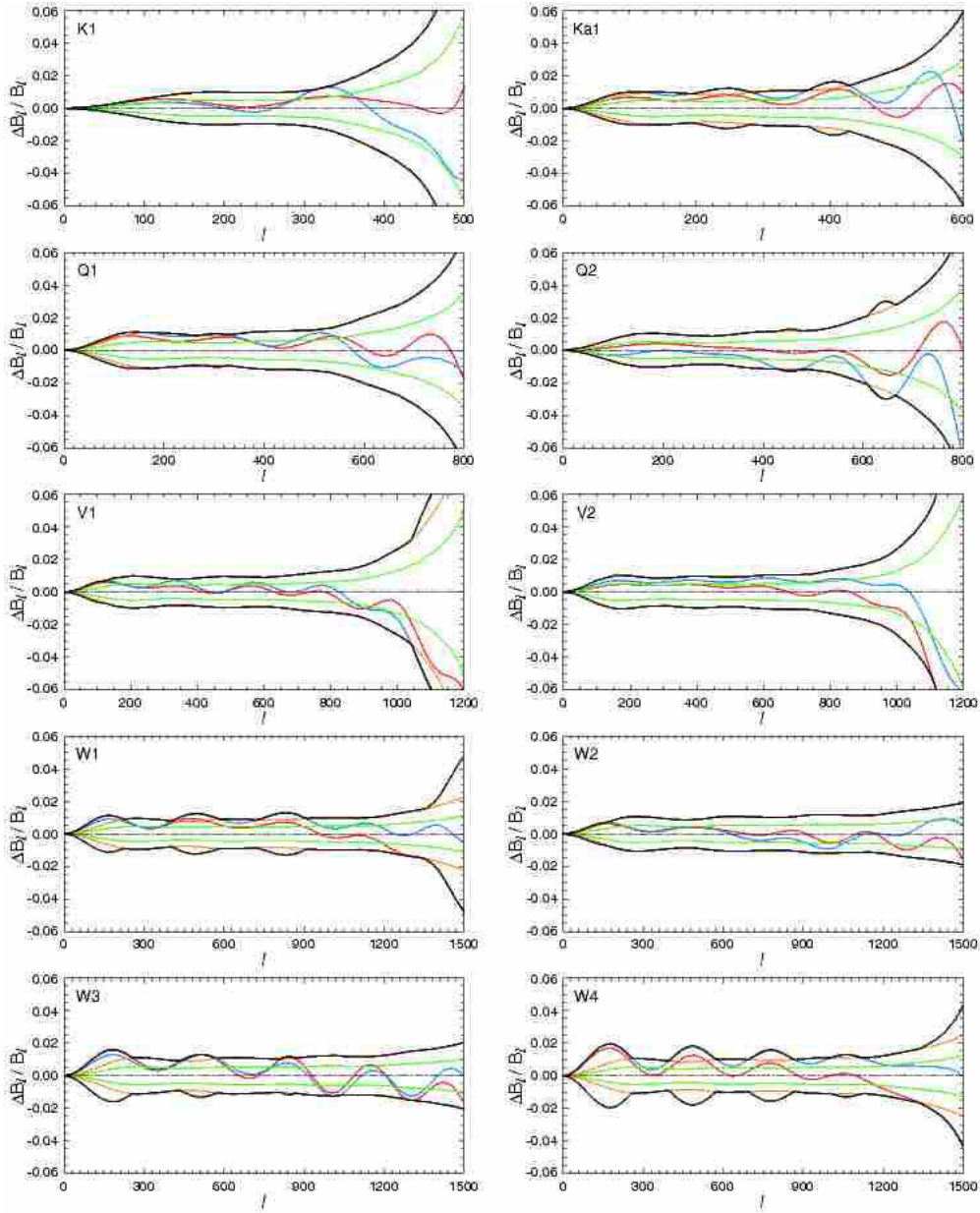}
\caption{The transfer functions and their statistical 
and systematic uncertainty.
The y axis of each panel shows the fractional uncertainty. The green curve
is the statistical uncertainty in the Hermite-based 
transfer function. The orange curve corresponds to 
twice the statistical uncertainty (it is mostly hidden by the black curve).  
The red curve is the fractional difference between the 
$b_l$ computed from the spherical
harmonic decomposition of the time stream and the Hermite fit. 
The blue line is the fractional difference between
the $b_l$ derived from the Juipter maps, after dividing by 
the $2.4^{\prime}$ pixelization window function, and the  
Hermite-based transfer function.
The black curve is the adopted 1$\sigma$ uncertainty used in all 
analyses. It corresponds to the absolute value of the 
maximum deviation from zero of the red, blue, and orange curves. 
The uncertainties 
on the window function, $w_\ell$, are twice these, but average down when 
multiple channels are combined. The uncertainty at $\ell=1$ is small because we
calibrate on the CMB dipole. The uncertainty in $\Omega_B$ is manifest at 
high $\ell$.}
\label{fig:winerrs}
\end{figure*}

Equation~\ref{eq:fll} gives the formal statistical uncertainty
in the transfer functions which is shown in Figure~\ref{fig:winerrs} 
for the ten DAs.

The solid angle is a scaling factor for the transfer functions, $b_\ell$,
and does not directly enter into the uncertainty of the window function.
However, the noise that leads to the uncertainty in $\Omega_B$
also produces the finite $C^{aa'}$. In particular, the uncertainty
in the $\ell$ cutoff in the transfer function is manifest as the increased
uncertainties at large $\ell$ in Figure~\ref{fig:winerrs}. 
Because we found that the
scatter in $\Omega_B$ was larger than that predicted by the noise
by a factor of two in W band (\S{\ref{sec:bu}} item 2), 
the statistical error bars from
equation~\ref{eq:fll} are inflated by a factor of two 
in all bands to account for current systematic uncertainties 
intrinsic to the Jupiter data.
This is shown in Figure~\ref{fig:winerrs}.

\subsection{Systematic Uncertainty in the Window Functions}

\label{sec:wluncert}
The window functions depend on the treatment of the beams, sky coverage, and
analysis method at the few percent level. These systematic effects 
are included in the window function uncertainty.

To determine the effect of incomplete symmetrization, mapping functions 
$M({\bf n},{\bf n'})$ are computed for coverage corresponding
to a pixel near the north ecliptic pole, which has the most 
symmetric coverage, and for a pixel on the ecliptic, which has the 
least symmetric coverage. For each mapping function, the full 
window function (equation~\ref{eq:w_yy}) is computed and compared to
$b_l^2$ computed from equation~\ref{eq:bl_hermite_exp}.  

\begin{figure*}[tb]
\plotone{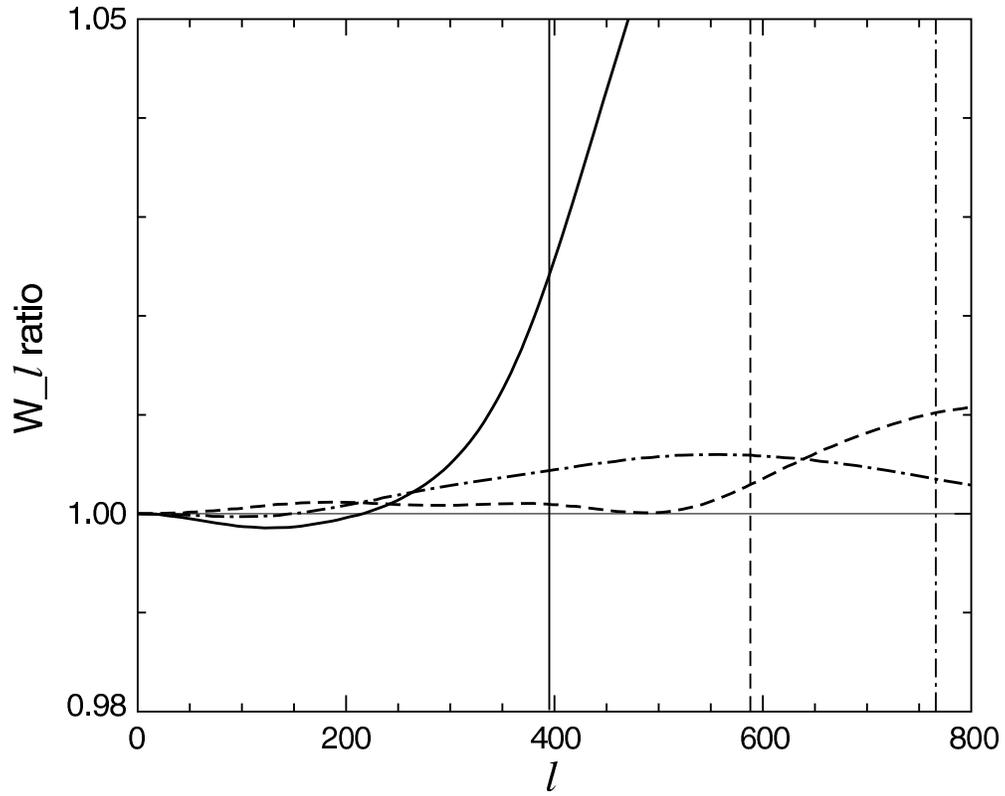}
\caption{A comparison of the window function of a fully 
symmetrized beam (characteristic of pixels near the ecliptic poles),
to the window function of a beam with only partial symmetrization
(characteristic of pixels near the ecliptic equator).
The solid line is for Q band, the dash line is for V band, 
and the dot-dashed line
is for W band. The vertical lines indicate where the 
window functions drop to 0.1 of their value at $\ell=0$.}
\label{fig:beam_asym}
\end{figure*}

Figure~\ref{fig:beam_asym} shows the  departure of the full window 
function from the
symmetrized window function as a function of $\ell$.
For all CMB analyses, the symmetric beam assumption is accurate to
1\% for all $\ell$ except for $300<\ell<400$ in Q band.
In this range, the instrument noise is significantly larger than 
the beam uncertainty. The uncertainty in the symmetrization 
is included in the year-one analysis
to the extent that it is accounted for in the uncertainties
shown in Figure~\ref{fig:winerrs}.

In addition to the Hermite-based method,
The transfer functions are computed from the Jupiter maps
( using equation~\ref{eq:bl} )
as well as from a direct transform of the binned TOD
(using equation~\ref{eq:w_yy}). These transforms weight the data 
differently, and less naturally, than do the Hermite expansion
and are thus considered as checks. 
The difference between the transform methods is shown in 
Figure~\ref{fig:winerrs}. As the 1~$\sigma$ error at each $\ell$, we adopt the 
maximum departure from zero of the two alternative transforms 
or twice the formal uncertainty derived from $C^{aa'}$.
This error bound is shown as the outer envelope 
in Figure~\ref{fig:winerrs} and is propagated through all other analyses. 

In \S{\ref{sec:bu}}, it was shown that the choice of 
$\theta_{Rc}$ leads to a possible
systematic bias in the determination of the total beam solid angle in W band.
The Jupiter-based window functions were computed with the extended  
$\theta_{Rc}$ (\S{\ref{sec:bu}}, item 3) and were found to differ 
from the baseline window functions 
by at most 
0.3\% for $\ell<\ell^{w=0.1}$. This demonstrates that $\approx 1$\% effects in 
Jupiter maps can be negligible in $\ell$-space.

\section{Beam Parameters for Coadded Data Sets}

Most analyses are performed on maps that have been 
coadded by polarization or frequency 
\citep{bennett/etal:2003b,hinshaw/etal:2003}.
Table~\ref{tab:fpa_beams5} gives the effective solid angles, gains
and frequencies for two map combinations. 
The effective solid angle here is computed with 
equation~\ref{eq:bs}, so it does
not include the $2.4^\prime$ Jupiter pixelization.  These values
should be used for data analyses of the sky maps.

The conversion from flux in Janskys ($10^{-26}$ Wm$^{-2}$Hz$^{-1}$)
to antenna temperature depends on the beam. The flux is 
modeled as $F_\nu \propto (\nu/\nu_e)^\alpha$. For a
broad band receiver for which the gain is known at all frequencies,
the conversion factor is:
\begin{equation}
\Gamma^{bb} = { (c^2 / 8\pi k_B\nu_e^2 )
\int f(\nu) G_m(\nu)(\nu/\nu_e)^{\alpha-2}d\nu  \over 
\int f(\nu) d\nu}
\label{eq:jansk}
\end{equation} 
where $G_m$ is the maximum gain, $f(\nu )$ is the radiometer band pass, 
and $k_B$ is Boltzmann's constant.
The ``bb'' superscript indicates that this is a broad band
quantity computed from a model of the optics. 

For {\WMAP} we report an effective conversion factor of 
$\Gamma=c^2/(2k_B\Omega^S(\nu_e^{ff})^2)$ for sources with $\alpha=-0.1$.
Here, $\Omega_B$ is computed from the Hermite beam profiles and
$\nu_e^{ff}$ is the effective frequency for free-free emission. 
The fractional uncertainty is
the same as for $\Omega^S$. The factors are tabulated in  
Table~\ref{tab:fpa_beams5}.

\section{Conclusions}
We have presented the characteristics of the beams in both
real space and in $\ell$-space and assessed the uncertainties
in both domains. In K, Ka, Q, V, and W bands the uncertainties 
in the beam solid angles
are given by 2.6\%, 1.2\%, 1.2\%, 1.1\%, and 2.1\%. The uncertainties
in the window functions are typically 3\% at most values of $\ell$. 
These values include systematic effects.

We have also presented the formalism in which the
beam uncertainties are propagated throughout the analysis.
The uncertainties which we adopt are conservative though prudent
for this stage of the data analysis. 

The K-band sidelobe and W-band pedestal corrections are 
the only beam-related effects that
are added in ``by hand'' and not treated in the formalism. These 
effects are significant for real space analyses and are 
currently negligible for the $\ell-$space CMB analyses.

Jupiter is mapped approximately twice per year. With more data
and improved modeling, our knowledge of the beams and window functions 
will improve over the length of the mission. 

The Jupiter maps and window functions are available
on-line through the LAMBDA web site at \url{http://lambda.gsfc.nasa.gov/}.

\acknowledgements

We thank Mike Nolta for useful discussions throughout the preparation of 
this paper. Consultations with YRS Associates (rahmat@ucla.edu) and 
their {\tt DADRA} code have
played a central role in the development and assessment of the {\WMAP}
optics.  Ken Hersey led the beam mapping effort at NASA/GSFC in addition to
working on the beam predictions. Our ability to test the beam model
is rooted in his work. Cliff Jackson at NASA/GSFC 
guided the {\WMAP} thermal reflector system through all phases 
of development. The success of the {\WMAP} optical system is largely
the result of his tremendous efforts. We are also grateful
for the dedicated work of many engineers 
and technicians who made {\WMAP} a reality.

\section{Appendix A}

The covariance matrix of the normalized beam is given by
\begin{equation}
\Sigma_{\ell \ell'}^b = \sum_{i,j=0}^{m_h}
{\partial b_\ell\over \partial a_{2i}} C^{aa'}_{ij}
{\partial b_{\ell'}\over\partial a_{2j}}
\label{eq:a1}
\end{equation}
where 
\begin{equation}
b_\ell = {B_\ell\over B_0} = {\sum_k a_{2k}B_{\ell k}\over \sum_k a_{2k}B_{0k}}
\end{equation}
and
\begin{equation}
{\partial b_\ell\over \partial a_{2i}} =
{B_{i\ell} \over  \sum_k a_{2k}B_{0k}} - 
{\sum_k a_{2k}B_{\ell k} \over (\sum_k a_{2k}B_{0k})^2}B_{0i} .
\label{eq:a3} 
\end{equation}
With the real space beam normalized to unity at $\theta=0$,
$\sum_k a_{2k}B_{0k}=\Omega^S$. After plugging 
equation~\ref{eq:a1} into equation~\ref{eq:a3} there are four terms similar
in form to
\begin{equation}
{\sum_k a_{2k}B_{\ell k} \over (\sum_k a_{2k}B_{0k})^2}B_{0i} C^{aa'}_{ij}
{\sum_j a_{2k}B_{\ell' k} \over (\sum_k a_{2k}B_{0'k})^2}B_{0'j}
= {b_\ell b_{\ell'}\Sigma_{00}\over(\Omega^S)^2}.
\end{equation}
After working through the the other terms, we find equation~\ref{eq:fll}.

\clearpage
\begin{deluxetable}{ll}
\tablecaption{Selected \WMAP\ beam analysis notation}
\tablewidth{0pt}
\tablehead{
\colhead{Symbol(s)} & \colhead{Description}}
\startdata
$T_J^m$                &  Measured amplitude of Jupiter \\
$T_J$                &  Brightness temperature of Jupiter \\
$\Omega_J^{ref}$       &  Reference solid angle of Jupiter \\
$\Omega_B$             & Generic main beam solid angle \\
$\Omega_B^A$ \& $\Omega_B^B$  & Main beam solid angle for A and B sides  \\
$\Omega_S$             & Generic sidelobe solid angle \\
$\Omega_A$         & Antenna solid angle, $\Omega_A=\Omega_B+\Omega_S$  \\
$\Omega^S$    & Effective solid angle of symmetrized A and B beams combined \\
$b_T(\theta)$ &Full sky beam profile normalized with $\int bd\Omega=\Omega_A$\\
$b^S(\theta)$  &Normalized and symmetrized beam profile\\
$b_\ell$  &Beam transfer function normalized to unity at $\ell=0$\\
$\Sigma^b_{\ell,\ell^\prime}$  &Beam covariance matrix for $b_\ell$\\
$B(\theta)$  &Beam profile normalized with $\int Bd\Omega=1$\\
$B_\ell$  &Beam transfer function of $B(\theta)$\\
$\Sigma^B_{\ell,\ell^\prime}$  &Beam covariance matrix for $B_\ell$\\
$w_\ell$  & Window function normalized to unity at $\ell=0$\\
\enddata
\tablecomments{Throughout this paper, small letters ({\it e.g.}, $b, b_\ell$) 
are used to denote dimensionless quantities normalized at $\theta=0$ 
or $\ell=0$, and capital letters ({\it e.g.}, $B, B_\ell$) are 
used to denote quantities normalized with $\Omega_B$.
The term ``transfer function'' refers to the Hermite or Legendre transform
of a beam. The term
``window function'' is reserved for the square of the transfer function.}
\label{tab:notation}
\end{deluxetable}

\clearpage

\begin{deluxetable}{ccccccc}
\tablewidth{5.5in}
\tabletypesize{\small}
\tablecaption{Main Beam Forward Gain Budget\label{tab:gain_budget}}
\tablehead{
\colhead{Gain} &
\colhead{K} &
\colhead{Ka} &
\colhead{Q} &
\colhead{V} &
\colhead{W1 \& W4} &
\colhead{W2 \& W3}
}
\startdata
Maximum optical design gain (dBi) & 47.1 & 49.8 & 51.7 & 55.5 & 58.6 & 59.3 \\
$e^{-k^2\sigma_z^2}$ & 0.99 & 0.97 &0.95 & 0.90 & 0.78 & 0.78\\
After scattering (dBi) & 47.0 & 49.7 & 51.5 & 55.1 & 57.5 & 58.2\\
After sampling (dBi) & 47.0 & 49.6 & 51.4 & 54.9 & 57.3 & 57.9\\
Measured in flight (dBi) & 47.3 & 49.4  & 51.4 & 54.6 & 57.7  & 57.6  \\
\enddata
\tablecomments{
The forward gain budget for representative beams. 
The first row gives the as-designed forward gain that would be achieved
with a stationary satellite with ideal reflectors. The next line 
gives the gain reduction factor from the Ruze formula 
(\cite{ruze:1966}, $k$ is the wave vector,
$\sigma_z$ is given in the text). The scattering reduces the gain to the values
given in the line labeled ``after scattering.'' The finite integration
time for each data 
sample results in a slight smearing of the beam, reducing it to the values
in the line labeled ``after sampling.'' For reference, 1\% corresponds to
0.043~dB.}
\end{deluxetable}

\begin{deluxetable}{lccccccccc}
\tablewidth{6.1in}
\tabletypesize{\small}
\tablecaption{Main Beam Solid Angles, Gains, \& Effective Frequencies\label{tab:fpa_beams}}
\tablehead{
\colhead{Beam}  & \colhead{$\Omega_B$}   & \colhead{$G_{m}$} &
\colhead{$\nu_e^{sync}$} & \colhead{$\nu_e^{ff}$} & \colhead{$\nu_e^{cmb}$} & 
\colhead{$P1\nu_e^{cmb}$} & \colhead{$P2\nu_e^{cmb}$} & \colhead{$\nu_e^{RJ}$} &
\colhead{$\nu_e^{dust}$}\\
   & (sr) & (dBi) & (GHz) & (GHz) & (GHz) 
   & (GHz) & (GHz)    & (GHz) & (GHz)        
}
\startdata
K1A &$2.44\times10^{-4}$ &47.1 & 22.42 & 22.50 & 22.76 & 22.36&23.18&22.77& 22.95\\
K1B &$2.36\times10^{-4} $ &47.3 & 22.47 & 22.55 & 22.80 &&& 22.80 & 22.98\\
Ka1A&$1.43\times10^{-4} $ &49.4 & 32.67 & 32.74 & 32.98 & 32.84&33.19&32.99& 33.17\\
Ka1B&$1.45\times10^{-4} $ &49.4 & 32.61 & 32.68 & 32.92 &&& 32.93 & 33.10\\
Q1A &$0.869\times10^{-4} $ &51.6 & 40.54 & 40.64 & 40.97 & 40.96&40.79&40.88& 41.18\\
Q1B &$0.903\times10^{-4} $ &51.4 & 40.58 & 40.67 & 40.94 &&& 40.99 & 41.20\\
Q2A &$0.911\times10^{-4} $ &51.4 & 40.48 & 40.56 & 40.83 & 40.84&40.23&40.99& 41.03\\
Q2B &$0.895\times10^{-4} $ &51.5 & 40.43 & 40.51 & 40.79 &&& 40.80 & 40.99\\
V1A &$0.405\times10^{-4} $ &54.9 & 60.24 & 60.40 & 60.91 & 59.32&61.22& 60.96&61.36\\
V1B &$0.439\times10^{-4} $ &54.6 & 60.22 & 60.38 & 60.90 &&& 60.95 & 61.35\\
V2A &$0.426\times10^{-4} $ &54.7 & 60.95 & 61.10 & 61.58 & 61.72&60.75& 61.63 & 62.00\\
V2B &$0.421\times10^{-4} $ &54.7 & 60.93 & 61.09 & 61.57 &&& 61.62 & 62.00\\
W1A &$0.196\times10^{-4} $ &58.0 & 92.85 & 93.08 & 93.72 & 93.71&93.27&93.89&94.47\\
W1B &$0.213\times10^{-4} $ &57.7 & 92.74 & 92.97 & 93.61 &&& 93.78 & 94.37\\
W2A &$0.214\times10^{-4} $ &57.7 & 93.33 & 93.50 & 93.97 & 93.58&94.35& 94.10&94.54\\
W2B &$0.233\times10^{-4} $ &57.3 & 93.38 & 93.55 & 93.04 &&& 94.17 & 94.61\\
W3A &$0.205\times10^{-4} $ &57.9 & 92.42 & 92.58 & 92.03 & 92.44&93.39& 93.15&93.57\\
W3B &$0.235\times10^{-4} $ &57.3 & 92.45 & 92.61 & 93.07 &&& 93.19 & 93.63\\
W4A &$0.203\times10^{-4} $ &57.9 & 93.25 & 93.45 & 94.01 & 94.33&93.18& 94.17&94.68\\
W4B &$0.213\times10^{-4} $ &57.7 & 93.19 & 93.39 & 93.96 &&& 94.11 & 94.63\\
\enddata
\tablecomments{The $\Omega_B$ are derived from the Jupiter maps and include
the smearing from the $2.4^\prime$ pixelization. These solid angles
should be used for working with the Jupiter maps.
The forward gain is $G_m=4\pi/\Omega_B$. The effective frequencies are
for sources smaller than the beam size (except for entries P1 and P2). 
For diffuse sources such as the CMB anisotropy, one should 
use the tabulation in \cite{jarosik/etal:2003}.
The columns with the P1 and P2 labels are for the two polarizations and come
directly from Table 11 in 
\cite{jarosik/etal:2003}. They are the same for the A and B sides.
By comparing P1 and P2 with $\nu_e^{cmb}$ from the previous column, 
one can assess the effects
of the optical gain on the passband. The $1\sigma$ uncertainty 
on the effective frequency is $0.1$~GHz
though the values are given to 0.01~GHz so the trends may be assessed.
}
\end{deluxetable}

\begin{deluxetable}{ccccccc}
\tablewidth{3.6in}
\tabletypesize{\small}
\tablecaption{Transfer Function and Window Function Parameters\label{tab:wins}}
\tablehead{
& \colhead{K} & \colhead{Ka} & \colhead{Q} & \colhead{V} &
\colhead{W$_{14}$} & \colhead{W$_{23}$}
}
\startdata
$m_h$ & 10 & 30 & 30 & 50/70 & 70 & 70 \\
$\sigma_h$ (deg) & 0.348 & 0.268 & 0.210 & 0.139 & 0.089 & 0.088 \\
$\ell^{w_\ell=0.5}$ & 128  & 163 & 200 & 265 & 300 & 260 \\
$\ell^{w_\ell=0.1}$ & 240 & 320 & 400 & 580 & 840 & 700 \\
\enddata
\tablecomments{
Parameters used in the beam fits and their assessment.  The number
of terms in the Hermite expansion is given by $m_h$ and the 
Gaussian width for the expansion are given by $\sigma_h$.
The $\ell$ value for which the normalized window is 0.5 is
$\ell^{w_\ell=0.5}$ and similarly for $\ell^{w_\ell=0.1}$. }
\end{deluxetable}

\begin{deluxetable}{cccccc}
\tablewidth{4in}
\tabletypesize{\small}
\tablecaption{Main Beam Solid Angles, Gains, $\Gamma$,
and Effective Frequencies for Combined Maps\label{tab:fpa_beams5}}
\tablehead{
\colhead{Beam}  & \colhead{$\Omega^S$} & \colhead{$\theta_{FWHM}$} &
\colhead{$G_{m}$} & \colhead{$\Gamma$} & \colhead{$\nu_e^{cmb}$} \\
      &    (sr)    &     (deg)       & (dBi)     & ($\mu$K/Jy)   & (GHz) 
}
\startdata
\sidehead{For 10 maps}
K   &$2.39\times 10^{-4}$  &0.82 & 47.2 & 268 & 22.8\\
Ka  &$1.43\times 10^{-4}  $  &0.62 & 49.4 & 213 & 33.0\\
Q1   &$0.879\times 10^{-4} $ &0.48 & 51.6 & 224 & 40.9\\
Q2   &$0.900\times 10^{-4} $ &0.48 & 51.4 & 220 & 40.5\\
V1   &$0.418\times 10^{-4} $ &0.33 & 54.8 & 214 & 60.3\\
V2   &$0.416\times 10^{-4} $ &0.33 & 54.8 & 210 & 61.2\\
W1   &$0.199\times 10^{-4} $ &0.21 & 58.0 & 190 & 93.5\\
W2   &$0.215\times 10^{-4} $ &0.20 & 57.7 & 173 & 94.0\\
W3   &$0.213\times 10^{-4} $ &0.20 & 57.7 & 179 & 92.9\\
W4   &$0.202\times 10^{-4} $ &0.21 & 57.9 & 185 & 93.8\\
\sidehead{For 5 maps}
K   &$2.39\times 10^{-4}  $ &0.82  & 47.2 & 269 & 22.8\\
Ka  &$1.43\times 10^{-4}  $ &0.62  & 49.4 & 213 & 33.0\\
Q   &$0.889\times 10^{-4} $ &0.49  & 51.5 & 222 & 40.7\\
V   &$0.417\times 10^{-4} $ &0.33  & 54.8 & 212 & 60.8\\
W   &$0.206\times 10^{-4} $ &0.21  & 57.8 & 182 & 93.5\\
\enddata
\tablecomments{
The values for $\nu_e^{cmb}$ are the average of the values in one band as given
by \protect{\cite{jarosik/etal:2003}}. They are appropriate for the 
CMB anisotropy. The top ten entries are for the ten
maps in which the two polarizations have been combined. The bottom five are for
the maps combined by polarization and band. A useful characteristic
beam resolution is the full width at half the beam maximum, $\theta_{FWHM}$,
though the beams are not Gaussian. The values for $\Gamma$ are for a source
with a free-free spectrum. The K-band value is appropriate for 
the sidelobe corrected map described 
in \protect{\cite{hinshaw/etal:2003}}.
For year-one analyses using $\Omega^S$ and $\Gamma$, we recommend
uncertainties of 2.6\%, 1.2\%, 1.2\%, 1.1\%, and 2.1\% in K through W band
respectively.}
\end{deluxetable}

\clearpage

\end{document}